\documentclass[a4paper,10pt,twoside]{cpc-hepnp}
\usepackage{CJK,upgreek,fancyhdr}
\usepackage{multicol}
\usepackage{graphicx}
\usepackage{booktabs}
\usepackage{amssymb,bm,mathrsfs,bbm,amscd}
\usepackage[tbtags]{amsmath}
\usepackage{lastpage}
\usepackage{multirow}
\usepackage{color}

\begin{document}
\begin{CJK*}{GB}{gbsn}

\fancyhead[c]{\small Chinese Physics C~~~Vol. 40, No. 11 (2016) 116001}
\fancyfoot[C]{\small 116001-\thepage}


\title{Alignment and measurement of the magnetic field for the BESIII muon counter\thanks{Supported by National Key Basic Research Program of China (2015CB856701); National Natural Science Foundation of China (NSFC) (11475187,11575198,11521505); 100 Talents Program of CAS (U-25).}}

\author{%
      Qing Gao (高清)$^{1;1)}$\email{gaoq@mail.ihep.ac.cn}%
\quad Jing-Zhi Zhang (张景芝)$^{1;2)}$\email{jingzhi@mail.ihep.ac.cn}%
\quad Chun-Hua Li (李春花)$^2$
\quad Jun-Hao Yin (殷俊昊)$^1$
}
\maketitle

\address{%
$^1$ Institute of High Energy Physics, Chinese Academy of Sciences, Beijing 100049, China\\
$^2$ University of Melbourne, VIC 3010, Australia\\
}

\footnotetext[0]{\hspace*{-3mm}\raisebox{0.3ex}{$\scriptstyle\copyright$}2016
Chinese Physical Society and the Institute of High Energy Physics
of the Chinese Academy of Sciences and the Institute
of Modern Physics of the Chinese Academy of Sciences and IOP Publishing Ltd}%

\begin{abstract}
  Based on cosmic ray events without a magnetic field taken with the BESIII detector during the summer shutdown of BEPCII in 2012 and di-muon events from a data sample taken at center-of-mass energy of 3.686 GeV in 2009, we compare the coordinates of hits registered in the BESIII muon counter with the expected interaction point extrapolated from
reconstructed tracks from the inner tracking system in the absence of a magnetic field.
By minimizing the difference, we align the muon counter with the inner tracking system.
Moreover, the strength of the magnetic field in the muon counter is measured for the first time
with di-muon events from data taken at a center-of-mass energy of 3.686~GeV.
After the alignment and the magnetic field strength measurement, the offsets in the reconstructed hit positions
for muon tracks are reduced, which improves the muon identification. The alignment and magnetic field strength measurement have been adopted in the latest version of the BESIII offline software system. This addition to the software reduces the systematic uncertainty for the physics analysis in cases where the muon counter information is used.
\end{abstract}



\begin{multicols}{2}

\section{Introduction}

The Beijing Spectrometer III (BESIII) detector~\cite{BESIII} is a general-purpose magnetic spectrometer
operating at the Beijing Electron-Positron Collider II (BEPCII)~\cite{BEPCII}. It is primarily used for
measurements of the inclusive/exclusive final states of electron and positron collisions and the study of properties of composite particles and interaction laws.
The muon counter (MUC) is the outermost sub-detector of BESIII, with a solid angle coverage of
89\% of 4$\pi$. It plays an important role in separating muons from other charged particles, especially charged pions.
Limited by the energy losses of particles with inner sub-detectors, the minimum momentum for a muon
that can reach the MUC is approximately 0.4~GeV/$c$. The MUC is designed to provide a position resolution
of 2~mm with a detection efficiency of 95\% and a noise level lower than 0.4~Hz/cm$^2$.

When a particle passes through the MUC, the readout strips of the resistive plate chambers (RPC),
will fire and provide the spatial coordinate. The coordinate of each readout strip is assigned according
to the design. However, the real position~\cite{liangyt_MUC} might deviate from the
design due to imperfections in the installation or due to strip distortion over time. For these reasons,
a regular alignment procedure for the BESIII MUC is required. For this, we compare the positions of fired strips of the MUC
with the expected interaction points of extrapolated tracks from inner tracker, the main drift chamber (MDC)~\cite{BESIII}. The MUC is placed in the magnetic
flux return yoke and the magnetic field strength is calculated based on the field properties
in the MDC and on electromagnetic laws using the software package {\sc Ansys}~\cite{ansys}.
To optimize such a calculation, a measurement of the field strength is desired.

In this paper, we introduce a method to align the MUC with the MDC, which is similar
 to that described in Ref.~\cite{wangz_MUC}. We also measure the magnetic field strength in the MUC using
 separate $\mu^+$ and $\mu^-$ tracks obtained from beam data.
The MUC geometry information within the BESIII Offline Software System (BOSS)~\cite{BOSS} has been
updated based on the alignment results.

\section{The muon counter and alignment method}\label{methods}

The BESIII MUC consists of three parts, the barrel and the two endcaps, as shown in Fig.~\ref{struct}. Each part is made of RPCs sandwiched by iron absorbers.
The barrel part is octagonal in shape. It is composed of eight segments with nine layers.
A layer of a segment is referred to as a box. Each endcap has four segments with eight layers.
There are 136 boxes in total. The numbering of the boxes (boxId) for the east endcap, the barrel and the
west endcap are [0,31], [32,103], [104,135], respectively. The order of boxId for each segment of each part is from the innermost layer to outer layers.
The relative coordinates of the strips in one box are fixed independently of the absolute
position of the box. Hence, we only need to align the MUC in terms of a box.
The thicknesses of the iron absorbers of the barrel from the inner part outwards are three absorbers of 3~cm, two of 4~cm,
three of 8~cm, and one of 15~cm. The coordinates of the boxes are described in three orthogonal directions, $x$, $y$ and $z$. The $z-$axis is along the $e^-$ beam and the $y-$axis is upright.
In the barrel, the strips for the odd RPC layers are oriented azimuthally and provide the $z-$coordinate,
while those in the even layers are oriented in the $z$ direction and provide the $x-y$ coordinate. In the endcaps,
the $x-$coordinate for odd layers and $y-$orientation for even layers are read out.

\begin{center}
\includegraphics[width=5cm]{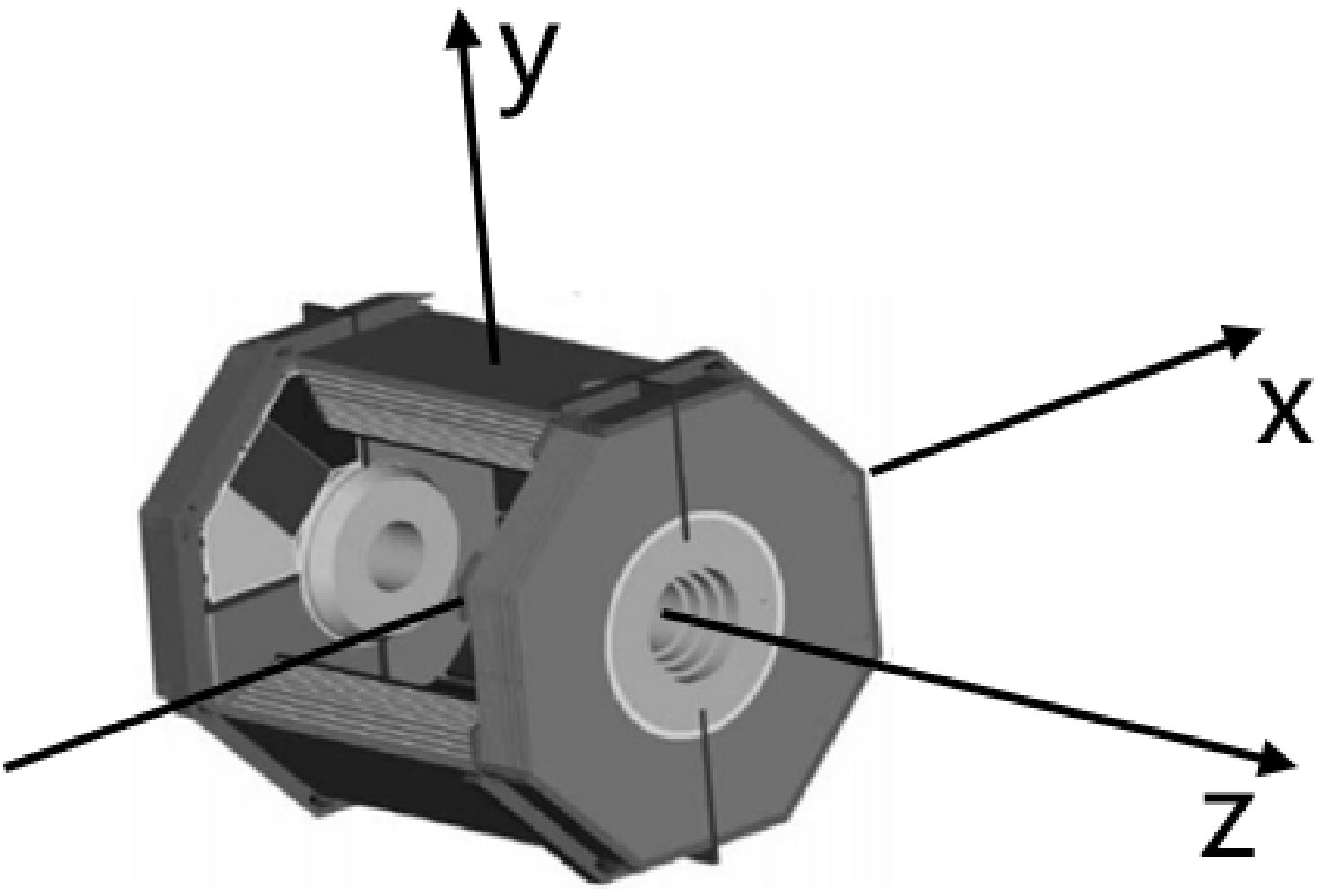}
\figcaption{\label{struct} Structure of the MUC and definition of the $x-,~ y-,~ z-$ directions.}
\end{center}

Without the magnetic field, we assume that a charged particle will keep its original direction and the
energy loss is due to the energy loss of ionization $dE/dx$. Muon decay and multiple scattering events are not excluded in the study.
We select muons from cosmic ray events taken during the summer shutdown of BEPCII in 2012 and compare the expected position with the measured position.
The expected position is the intersection of the muon track with each layer of the MUC. The muon track
is extrapolated from tracks reconstructed from the MDC to the RPCs with
{\sc Geant4}~\cite{GEANT4}. The measured position corresponds to the coordinate of the fired strip.
For well-aligned detectors, the two positions should coincide with each other. The difference between them,
i.e., the residual ($R = |\overrightarrow{V}_{\rm expect}-\overrightarrow{V}_{\rm obs}|$), comes from a misalignment of the MUC with the MDC. The residual of the strips in one box follow a Gaussian distribution, whose mean value represents the deviation between the two positions.
The standard deviation ($\sigma$) of the Gaussian represents the uncertainty of the extrapolated position.

The track of a charged particle bends in the magnetic field and the direction of curvature will depend upon the sign of its charge.
Any overestimate or underestimate of the field will result in a deviation of the particle
trajectory from its expectation. The sign of the deviation from the expected position for
$\mu^+$ and $\mu^-$ tracks is opposite. We use di-muon events produced in electron-positron collisions,
$e^+e^-\to\mu^+\mu^-$, to estimate the field strength.
By the residual of $\mu^+$ and $\mu^-$, we can measure the magnetic field strength.

\section{Alignment}

\subsection{Barrel}

A sample of cosmic ray events taken by BESIII without a magnetic field is used to align the barrel of the
MUC with the MDC. During this data taking, the endcaps of the MUC were not included, hence,
only the barrel part can be aligned by this data sample. When the muons of cosmic ray events pass through the
detector, they will leave straight tracks from the outermost layer of MUC to the inside of the detector and
leave in the opposite direction. To make use of the extrapolated tracks from the MDC, we select cosmic ray
events in which the muon passes through the MDC near the $e^+e^-$ interaction point (IP).
Each of those events is reconstructed to have two tracks originating from the MDC.
The time difference between the two tracks, obtained from the information of the time-of-flight (TOF) detector~\cite{BESIII}, peaks around 6~ns. The selection criteria are discussed below.

In this analysis, only tracks in the barrel which have a polar angle of $|\cos\theta|<0.8$ are accepted.
Candidate events must have two charged tracks. The two tracks are back-to-back and the azimuthal and polar angles are required
to be $|\phi_1-\phi_2-\pi| <$ 0.6, $|\theta_1+\theta_2-\pi| <0.7$, respectively. The timing information
of the TOF satisfies $t1\in [-5,1.9]$ ns and $|\delta t|=|t1-t2|>$ 5 ns, where $t1$ and $t2$ are the timings of
the charged tracks.

After imposing the previously discussed selection criteria, we obtain a sample of muon
candidates in which the magnetic field was turned off. The sample is split into two parts. The first part is used for the alignment, and the other part is for validation. By plotting the residual of hits of each box for
the selected candidates and fitting the distribution, we get the averaged residual of every box in the barrel,
as shown by the red dots in Fig.~\ref{alignment} (boxId from 32 to 103). The uncertainty of the residual is statistical and very small. The residual obviously deviates
from zero for most boxes. We modify the coordinate of each box by adding the residual to the initial
coordinate and validate the effect by the second part of the sample. The blue stars in Fig.~\ref{alignment} (boxId from 32 to 103)
show the residuals of each box after alignment. The residuals are significantly reduced.

\end{multicols}
\ruleup
\begin{center}
\includegraphics[width=15cm]{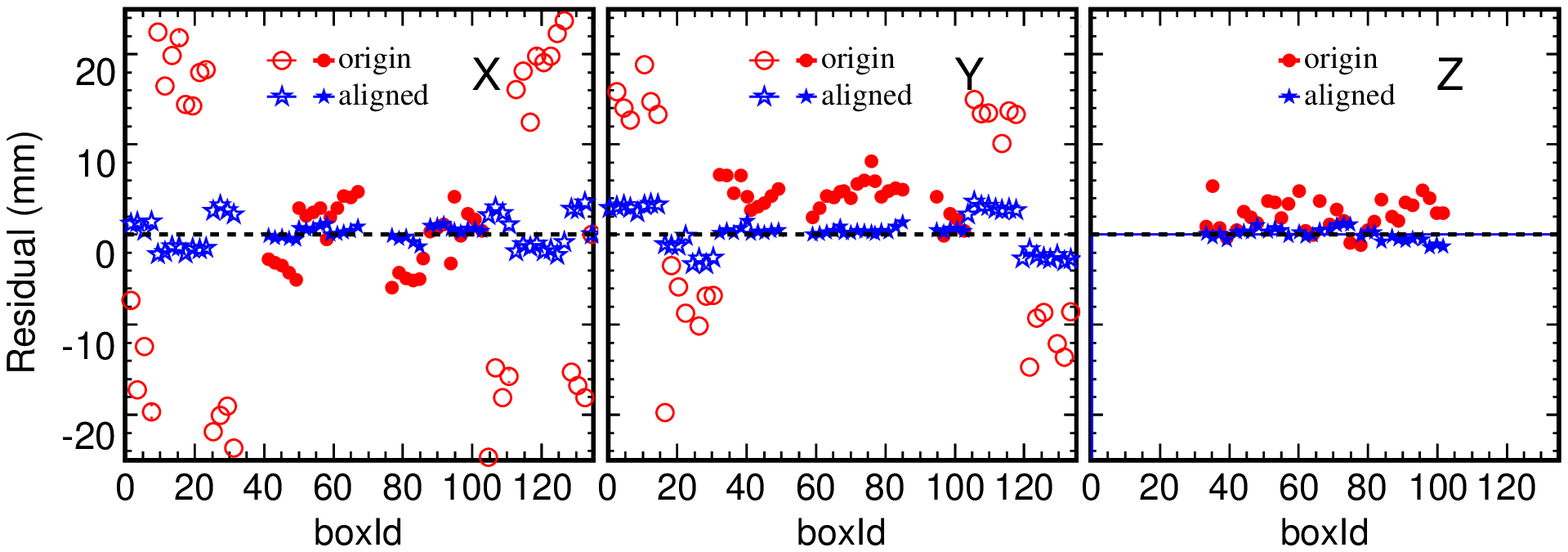}
\figcaption{\label{alignment} Residuals in $x-,~ y-,~ z-$ directions. The dashed line is the reference at zero, red circles and dots are before alignment, blue stars are after alignment. The barrel is aligned by cosmic ray events (solid markers), the endcaps (hollow markers) are aligned by di-muon events.}
\end{center}
\ruledown

\begin{multicols}{2}

\subsection{Endcaps}

To align the two endcaps, a di-muon sample selected from
data taken at 3.686 GeV is used, where the BESIII regular magnetic field was turned on. We do not use
cosmic ray events for the endcap alignment because of the low counting rate. Since the magnetic field will
bend $\mu^+$ and $\mu^-$ tracks in opposite directions, the residual caused by incorrect magnetic field values
should cancel if we sum $\mu^+$ and $\mu^-$, so the averaged residual of $\mu^+$ and $\mu^-$ indicates a misalignment of the detector.

To select di-muon candidates, charged tracks are required to originate from the IP
with $V_{xy}=\sqrt{V_{x}^{2}+V_{y}^{2}}<1\rm~cm$ and $|V_{z}|<10\rm~cm$,
where $V_{x}$, $V_{y}$, $V_{z}$ are the $x, y, z$ coordinates of the point of closest approach to the
run-dependent IP, respectively. Only tracks in the endcaps are accepted with polar angles in the
region $0.8<|\cos\theta_{1,2}|<0.93$. The momentum ($p$) of each track should be larger than 1.7~GeV/$c$.
The ratio of energy deposit ($E$) in the electromagnetic calorimeter (EMC)~\cite{BESIII} to the reconstructed momentum is required to be within
$0.06<E/p<0.2$ to suppress backgrounds from Bhabha events. Candidate events are required to have
two oppositely charged tracks and the two tracks are back-to-back with $px_{\rm tot} <$ 0.1 GeV/$c$,
$py_{\rm tot} <$ 0.1 GeV/$c$, $pz_{\rm tot} <$ 0.1 GeV/$c$, the angles between the two tracks
$|\phi_1-\phi_2-\pi| <$ 0.4, $|\theta_1+\theta_2-\pi| <$ 0.05, where $\phi_{1,2}$ and $\theta_{1,2}$
are the azimuthal and polar angles of the two tracks, respectively. The timing information of the TOF
satisfies $|\delta t|=|t1-t2| <$ 4~ns to suppress cosmic ray events, where $t1$ and $t2$ are the timing of the  TOF for the charged tracks.

After imposing the previously described selection criteria, we obtain a di-muon sample. This sample is also split into two parts. The first part is used for the alignment, and the other part is for validation. By the same method as used for the barrel analysis, we get the residual of fired hits for
selected candidates for each box in the endcaps, as shown by the red circles in Fig.~\ref{alignment}
(boxId in range $[0,31]\cup[104,135]$). The uncertainty of the residual is statistical and very small. The residual obviously deviates from zero for most boxes.
The coordinate of each box is modified by the same method as was used for the barrel and we validate the effect by the second part of the sample. The empty blue  stars in Fig.~\ref{alignment} (boxId in range $[0,31]\cup[104,135]$) show that the residual of each box after the alignment is much closer to zero.

\section{Magnetic field strength measurement}

Since there are no large changes in the residuals in the $z-$direction of
$\mu^+$ and $\mu^-$ tracks after alignment, implying that the magnetic field in the $r-\phi$ direction is
simulated well by {\sc Ansys}, we only focus on the measurement of the magnetic field in the $z$-direction
of the barrel.
By using the residuals in the $r-\phi$ direction for $\mu^+$ and $\mu^-$ tracks separately, the magnetic field strength
can be measured. The correct strength of the field is the current value (based on the calculation using {\sc Ansys}) multiplied by a factor that is determined by minimizing the residuals. We will get 72 factors (8 segments $\times$ 9 layers) for the magnetic field strength measurement.

\end{multicols}
\ruleup
\begin{center}
\includegraphics[width=7.5cm]{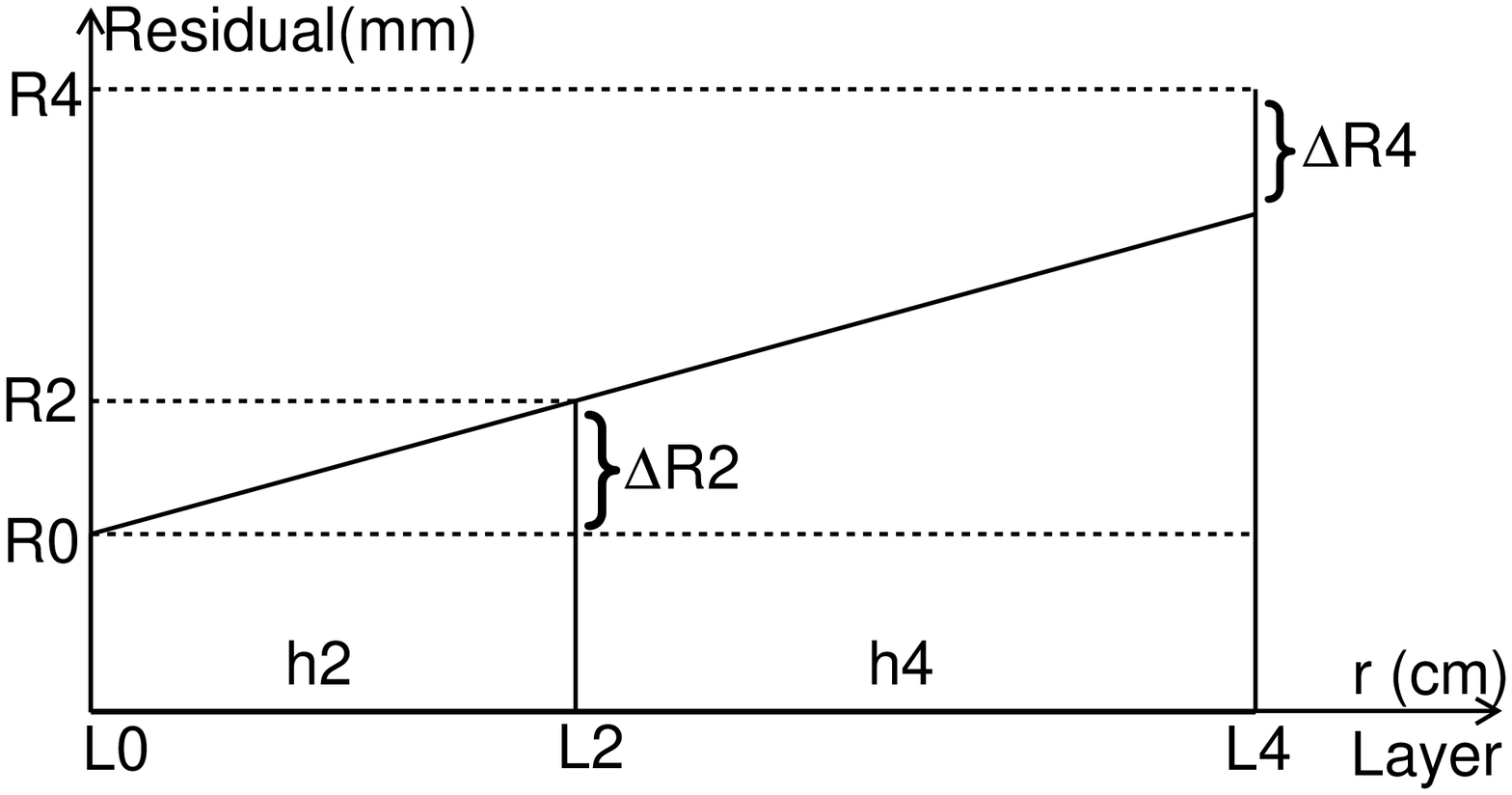}
\includegraphics[width=5cm]{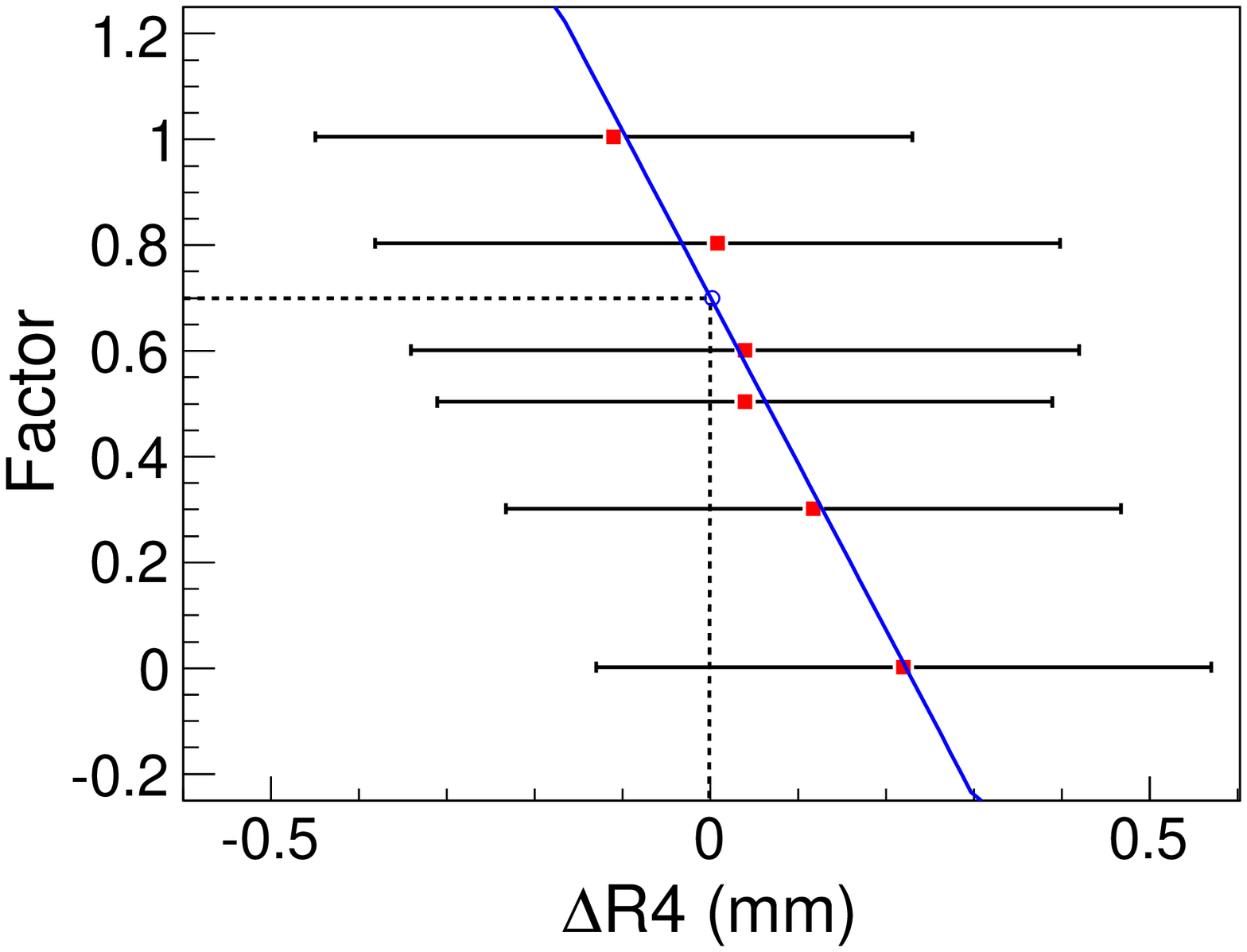}
\figcaption{\label{correct} Left, the residuals of layer 0, 2, 4 and the increase of residual $\Delta R2$, $\Delta R4$. Right, by fitting to the factors as described in the text, we obtain the optimum solution.}
\end{center}
\ruledown

\begin{multicols}{2}

The left-hand panel in Fig.~\ref{correct} illustrates the accumulation of residuals.
Only the even layers are shown because the
odd layers give only the $z-$direction information. The variables $L0,~L2,~L4$ refer to the
layers 0, 2, and 4, respectively; $h2 = 14$ cm is the distance from $L0$ to $L2$ and $h4 = 15$ cm is that
from $L2$ to $L4$; $R0,~R2,~R4$ are the residuals in $L0$, $L2$, $L4$, respectively.
$R0$ mainly comes from the statistical uncertainty of alignment for layer 0. $R2$ is partially from $R0$,
and partially from $\Delta R2=R2-R0$ due to the inaccuracy of magnetic field of $L0$. The residual $R4$
comes from three parts. The first is $R0$. The second part comes from the magnetic field inaccuracy
of $L0$. By assuming that the residual increases linearly along the trajectory of the particle,
the contribution of this part is $\frac{h2+h4}{h2}\times\Delta R2$. The third part is the pure contribution
coming from the magnetic field inaccuracy of the former layer,
$\Delta R4=R4-R0-\frac{h2+h4}{h2}\times\Delta R2$. $\Delta R2$ and $\Delta R4$ are used to calculate
the magnetic field of $L0$ and $L2$. Similarly, we can obtain $\Delta R6$ and $\Delta R8$, the pure contribution of residual for layer 6 ($L6$) and layer 8 ($L8$) coming from the magnetic field inaccuracy of the $L4$ and $L6$, which are used to get the magnetic field for $L4$ and $L6$ and are independent of the magnetic field for former layers. Thus we can get the strength of the magnetic field for all layers at the same time.

We vary the strength of magnetic field in the MUC by multiplying by a factor. Subsequently, the data are reconstructed
with the modified settings and we obtain new residuals. This procedure is repeated several times
by randomly selecting factors \{0,0.3,0.5,0.6,0.8,1\}.
As an example, the right-hand panel of Fig.~\ref{correct} shows the scatter plot of the correction factor versus
$\Delta R4$ for one segment. We fit the relation by a straight line to get the factor that is
associated with $\Delta R4$ equal to zero. Using the same method, we obtained the factors for the
other boxes as well. For the outermost layer ($L8$), no correction is used because
we do not have much information available to measure the magnetic field strength.
By assuming that the magnetic changes linearly, the odd layers take the average of their two neighboring
layers. The results of the measurements for each box are listed in Table~\ref{mag_factor}, where the uncertainty of the factor is statistics-dominated and about 0.03, which is ignored.
The factors for $L0$ are around zero and sometimes negative because the $L0$ RPC is in the middle of
the superconducting solenoid and iron absorber. The magnetic field strength in the $L0$ RPC is about $10^{-4}$ T,
which is much weaker than that in absorber iron. {\sc Ansys} cannot describe well such small magnetic field
strength and the very small correction factors in $L0$ can be accepted. The factors becomes larger in the
outer layers, which implies that the drop in the magnetic field strength is slower than given by {\sc Ansys}.

The results after reconstructing the data with the updated magnetic field strength are shown in Fig.~\ref{2-old}.
Clearly, the residuals of $\mu^{+}$ and $\mu^{-}$ tracks for selected di-muon events are reduced.

\end{multicols}
\begin{center}
\tabcaption{\label{mag_factor} The magnetic correction factors for each box.}
\footnotesize
\begin{tabular*}{170mm}{@{\extracolsep{\fill}}c|llllllllll}
\toprule
Segment&\multirow{1}{0.2in}{}&\multirow{1}{0.5in}{$L0$}&\multirow{1}{0.5in}{$L1$}&\multirow{1}{0.5in}{$L2$}&\multirow{1}{0.5in}{$L3$}&
\multirow{1}{0.5in}{$L4$}&\multirow{1}{0.5in}{$L5$}&\multirow{1}{0.5in}{$L6$}&\multirow{1}{0.5in}{$L7$}&
\multirow{1}{0.5in}{$L8$}\\
\hline
0            &&0.037&0.33&0.62&0.7 &0.89&1.44&2.00&1.50&1 \\
1            &&0.089&0.59&1.09&0.94&0.78&1.39&2.00&1.50&1 \\
2            &&-0.26&0.48&1.20&1.10&1.00&1.26&1.52&1.26&1  \\
3            &&0.07 &0.58&1.10&1.03&0.96&1.18&1.40&1.20&1 \\
4            &&-0.28&0.21&0.70&1.23&1.76&1.41&1.08&1.04&1 \\
5            &&0.24 &0.51&0.78&0.69&0.60&1.30&2.00&1.50&1    \\
6            &&-0.27&0.32&0.90&0.92&0.95&1.50&2.06&1.53&1\\
7            &&0.09 &0.49&0.89&0.82&0.76&1.41&2.06&1.53&1\\
\bottomrule
\end{tabular*}%
\end{center}

\begin{multicols}{2}

\begin{center}
\includegraphics[width=7.5cm]{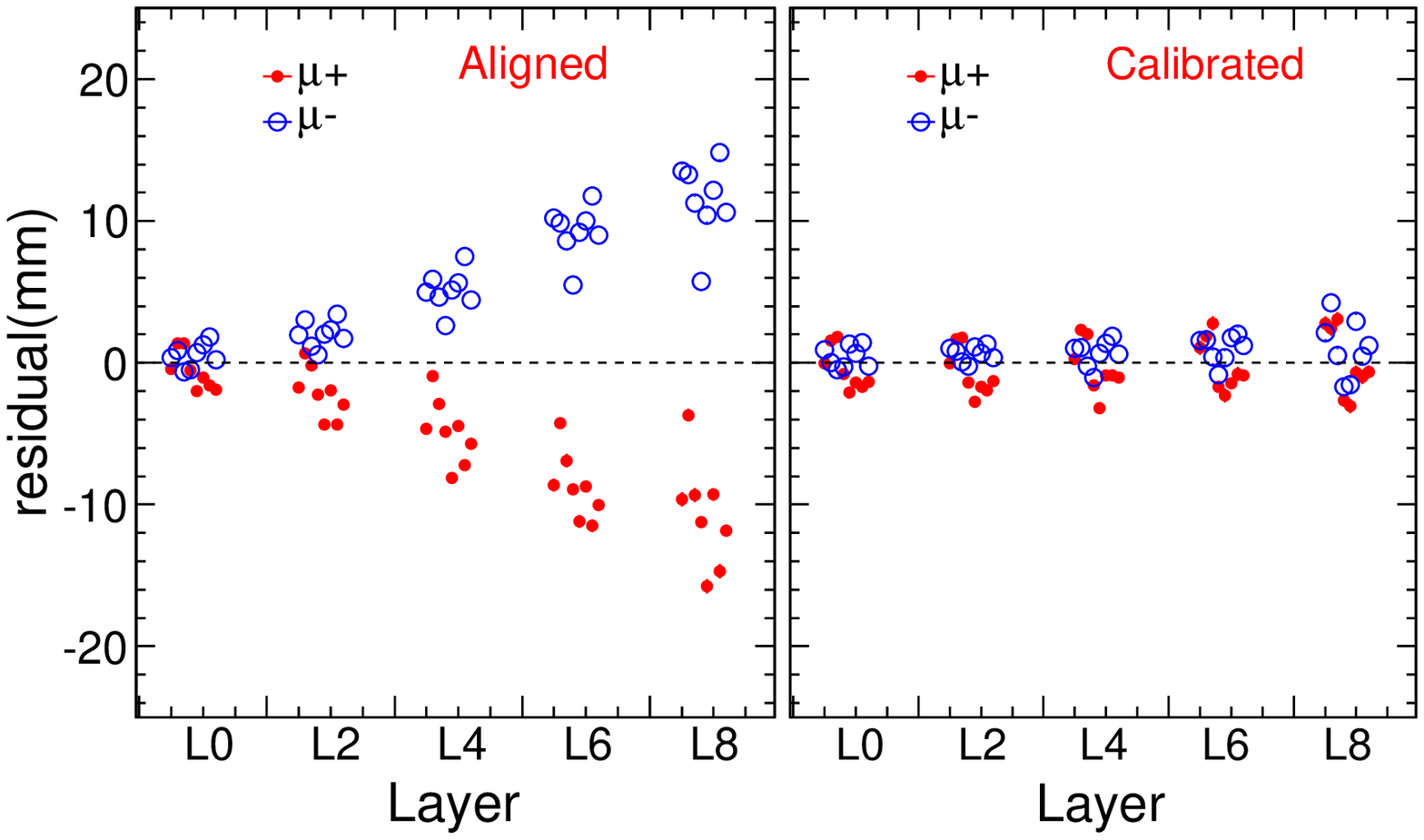}
\figcaption{\label{2-old} Measurement of the magnetic field strength. The residuals of reconstructed data before (left) and after (right) measurement of the field. Blue circles are $\mu^{-}$, red dots are $\mu^{+}$, and the dashed line is the reference line at zero.}
\end{center}
\ruledown

\section{Validation of the alignment and magnetic field strength measurement with $\gamma_{\rm ISR}\mu^{+}\mu^{-}$}

To validate the alignment and magnetic field strength measurement, we select 0.2 million $\gamma_{\rm ISR}\mu^{+}\mu^{-}$ events from a data sample collected at a center-of-mass energy corresponding to the mass of the $J/\psi$.
The residuals before the alignment and magnetic field strength measurements of $\mu^+$ and $\mu^-$ tracks are
shown in the left-hand panel of Fig.~\ref{gamamumu}. Apart from an additional requirement of $0.1 < E/p < 0.4$,
the selection criteria of charged tracks are the same as were used for the selection of di-muon events.
Candidates are required to have at least one photon candidate with deposited energy in the
EMC larger than 40~MeV.
A four-constraint (4C) kinematic fit is performed to the two tracks and the photon candidate with respect
to the initial $J/\psi$ four-momentum. A goodness-of-fit of $\chi^2_{\rm 4C} < 15$ is required.
The track with the largest momentum is required to have a depth (the distance that the charged particle travels) in the MUC larger than 41 $\rm{cm}$ and more than
4 layers of RPCs should be fired. The momentum of each track is required to be in the range (1, 1.5)~GeV/$c$
to reject the events from $J/\psi\to\mu^+\mu^-$, which tend to have a momentum around 1.55 GeV/$c$.
The residuals for $\mu^+$ and $\mu^-$ before the alignment deviate from zero and become larger
for the outer layers. After the alignment and the correction of the magnetic field strength, we obtain the new residuals
for $\mu^+$ and $\mu^-$ tracks. These results are shown in the right-hand panel of
Fig.~\ref{gamamumu}. The residuals are around zero after the corrections.

\ruleup
\begin{center}
\includegraphics[width=7.5cm]{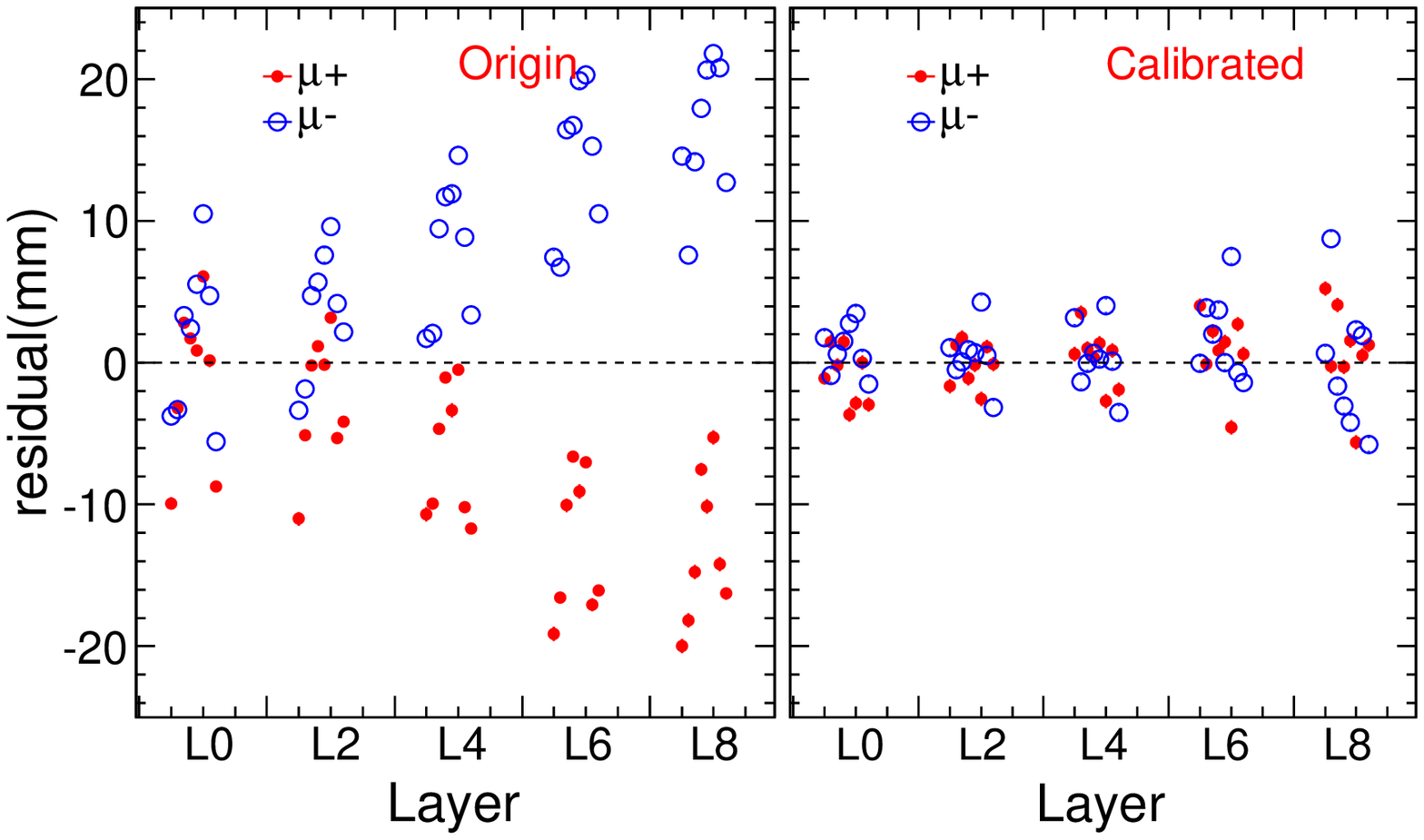}
\figcaption{\label{gamamumu} The residuals for selected $\gamma_{\rm ISR}\mu^{+}\mu^{-}$ samples before (left) and after (right) the alignment and the magnetic field strength measurement. Blue circles are $\mu^{-}$, red dots are $\mu^{+}$, and the dashed line is the reference line at zero.}
\end{center}
\ruledown


\section{Conclusion}

Based on the cosmic ray events taken without a magnetic field and di-muon events selected from data
taken at a center-of-mass energy of 3.686~GeV with the magnetic field turned on, we successfully aligned the BESIII
MUC with the inner tracking system (MDC). With the di-muon events, we also measured the magnetic field strength in the MUC.
After the alignment and the update of the magnetic field strength, the offsets in the hit positions
of $\mu^+$ and $\mu^-$ tracks diminish, which improves the muon identification. The work presented in the paper will improve BESIII data analysis involving the information from the MUC.

\vspace{-1mm}
\centerline{\rule{80mm}{0.1pt}}

\end{multicols}

\clearpage
\end{CJK*}
\end{document}